\documentclass[aps,prb,10pt, twocolumn, superscriptaddress, longbibliography, nobalancelastpage]{revtex4-1}

\usepackage{graphicx}
\usepackage{graphics}
\usepackage{amsmath}
\usepackage{amssymb}
\usepackage{amsfonts}
\usepackage{dsfont}
\usepackage{braket}
\usepackage{color}
\usepackage{braket,slashed}
\usepackage[mathscr]{euscript}
\definecolor{darkblue}{rgb}{0, 0, 0.8}
\usepackage[colorlinks=true, breaklinks=true, linkcolor=red, citecolor=blue, urlcolor=blue]{hyperref} 
\usepackage{hyperref}
\usepackage{subfigure}
\usepackage{xfrac}
\usepackage{bm}
\usepackage{kantlipsum}
\usepackage{enumitem}
\usepackage{tikz}
\usepackage{framed}
\usepackage{graphicx}
\usepackage{subfigure}

\allowdisplaybreaks[1]

\newcommand{\code}[1]{\texttt{#1}}

\newcommand{\one}{\mathds{1}}

\newcommand{\e}{\ensuremath{\mathrm{e}}}

\newcommand{\diagram}[2]{\;\vcenter{\hbox{\includegraphics[scale=0.32,page=#2]{./Diagrams/#1.pdf}}}\;}

\newcommand{\cluster}{cluster }

\newcommand{\pepo}{PEPO }

\begin{document}

\title{Symmetric \cluster expansions with tensor networks}

\author{Bram Vanhecke}
\email{bavhecke.Vanhecke@UGent.be}
\affiliation{Department of Physics and Astronomy, University of Ghent, Krijgslaan 281, 9000 Gent, Belgium}

\author{Laurens Vanderstraeten}
\affiliation{Department of Physics and Astronomy, University of Ghent, Krijgslaan 281, 9000 Gent, Belgium}

\author{Frank Verstraete}
\affiliation{Department of Physics and Astronomy, University of Ghent, Krijgslaan 281, 9000 Gent, Belgium}

\begin{abstract}
Cluster expansions for the exponential of local operators are constructed using tensor networks. In contrast to other approaches, the cluster expansion does not break any spatial or internal symmetries and exhibits a very favourable prefactor to the error scaling versus bond dimension. This is illustrated by time evolving a matrix product state using very large time steps, and by constructing a novel robust algorithm for finding ground states of 2-dimensional Hamiltonians using projected entangled pair states as fixed points of 2-dimensional transfer matrices.
\end{abstract}

\maketitle

The Trotter-Suzuki expansion \cite{Trotter1959, Suzuki1976} plays a ubiquitous role in theoretical physics as it relates quantum-mechanical problems in $d$ spatial dimensions to statistical-mechanical problems in $d+1$ dimensions. It is also a preferred tool for simulating the time-dependent Schr\"{o}dinger equation on a computer, as it allows to split a continuous differential equation into smaller steps $dt$ that can be integrated exactly; the corresponding error scales as $c_n.dt^{n+1}$ for an $n$'th order expansion, with $c_n$ a prefactor that can be optimized \cite{Barthel2019}. This Trotter-Suzuki expansion forms the basis for the time-evolving block decimation algorithms \cite{Vidal2003} for simulating quantum spin chains, and has been used extensively in that context for finding ground states (imaginary-time evolution) and for simulating real-time dynamics. The success of this method follows to a large extent from the fact that Trotter-Suzuki expansions are size extensive: they scale well in the thermodynamic limit. A very important drawback however is the fact that they always break spatial and/or internal symmetries of the underlying quantum system.  A different approach which preserves the symmetries of the Hamiltonian is the cluster expansion, widely used in the form of series-expansion methods in quantum Monte Carlo \cite{sandvik2019stochastic}. Such methods are however not size extensive, as they involve a power series of the Hamiltonian and can hence not be used for simulating uniform dynamics in the thermodynamic limit.
\par In this paper, we demonstrate that tensor-network techniques \cite{Verstraete2008} allow us to combine the advantages of both of those approaches into a practical and simple framework which is both size extensive and preserves global and spatial symmetries. The corresponding non-perturbative cluster expansion works equally well for simulating dynamics in one dimension with matrix product states \cite{Schollwoeck2011} (MPS) and in two dimensions with projected entangled-pair states \cite{Verstraete2004} (PEPS). Our expansion of the time-evolution operator can be understood as a sum of all possible clusters up to a certain size, but in such a way that the evolution in every individual cluster is exact up to infinite order. If we consider a case with nearest-neighbour interactions and for which the maximum cluster size is $p$, then only a fraction $p!/p^p\simeq \exp(-p)$ of the Hamiltonian terms in the Taylor expansion up to order $p$ will not be taken into account, leading to a very favourable error scaling of the prefactor $c_p$.
\par Our method can be seen as an efficient and practical way of implementing the ideas put forward in Refs.~\onlinecite{Hastings2006, Kliesch2014, Molnar2015}, where it was used to prove the polynomial scaling of the bond dimension of tensor-network algorithms for simulating quantum spin systems. It can also be understood as a vastly improved and practical version of the construction in Ref.~\onlinecite{Vanderstraeten2017}, in which PEPS wavefunctions were designed that bridge size-extensive perturbative expansions across a phase transition.

\renewcommand{\diagram}[1]{\,\vcenter{\hbox{\includegraphics[scale=0.3,page=#1]{./diagrams_mpo.pdf}}}\,}
\noindent\emph{Construction in one dimension.}---%
For illustrating our method we first consider the simplest case of a spin chain with nearest-neighbour interactions. We will construct a matrix-product operator (MPO) providing a cluster expansion for the exponentiated hamiltonian in the thermodynamic limit, i.e.
\begin{equation*}
\exp \left( t \sum_i h_{i,i+1} \right) \approx \diagram{1}.
\end{equation*}
The idea is to grow a matrix product operator which captures more and more terms in the perturbative expansion, but in such a way that the results in a given cluster are exact if Hamiltonian terms outside of the cluster are ignored.
\par The first non-zero element in the MPO tensor encodes the action of the unit operator, i.e. we set
\begin{equation*}
\diagram{2} = \one ,
\end{equation*}
where the label `$0$' denotes the first entry of the virtual legs. The first non-trivial cluster is of size two, and is obtained by exponentiating the nearest-neighbour interaction term and subtracting the unit that we have already incorporated. This cluster is encoded in the MPO by introducing a new virtual level in the MPO denoted by label `$1$' and with degeneracy depending on the Schmidt decomposition of the exponential of the local Hamiltonian under consideration. By performing a singular-value decomposition of this exponential minus the identity operator,  we get two new tensor elements $O'_{01}$ and $O'_{10}$
\begin{equation*}
\diagram{5} = \diagram{3} - \diagram{4},
\end{equation*}
We add these elements to the MPO, so $O\to O+O'$. Note that this cluster contains two-site terms of all orders and the extensivity property of the ensuing MPO is of crucial importance: the MPO contains an extensive number of terms with multiple non-overlapping clusters such as
\begin{equation*}
\diagram{6}.
\end{equation*}
\par The MPO does not contain overlapping terms, however, and therefore we have to introduce three-cluster terms. We obtain these by exponentiating the terms of the hamiltonian that act non-trivially on three sites, and subtract what was already contained in the two-cluster terms. Again, we can introduce a new element in the MPO tensor for capturing these terms
\begin{multline*}
\diagram{7} \\ = \diagram{8} - \diagram{9},
\end{multline*}
where on the right-hand side we sum over the unlabeled legs. We compute this tensor element $O'_{22}$ by applying the inverses of the $O_{1,0}$ and $O_{0,1}$ elements to the right hand side, and as such we do not have to increase the bond dimension of the MPO to achieve this. Adding this tensor element to the MPO, $O\to O+O'$ then encodes an extensive number of two- and three-site cluster terms.
\par We can continue this procedure, and incorporate four- and five-site clusters with an extra virtual level `$2$' and three new tensor elements $O'_{12}$, $O'_{21}$ and, in a next step, $O'_{22}$:
\begin{equation*}
\diagram{10} , \quad \diagram{11}.
\end{equation*}
\par This construction provides us with a MPO representation for cluster expansion of the exponentiated Hamiltonian, which conserves all spatial symmetries such as translation invariance and reflection symmetry. It is correct up to order $t^{p-1}$, where $p$ is the largest cluster size that is incorporated in the MPO, but, in fact, it captures a large portion of the higher-order terms as well. The terms of order $t^p$ that are lacking from the MPO representation are, indeed, only the $p+1$ clusters. There are $p!$ such terms in the exponential of size $p$, whereas we can count there are $p^p$ different $O(t^{p})$ terms with those same Hamiltonian terms. We thus find that the MPO gets a fraction $p!/p^p$ of the $O(t^{p})$ terms wrong, or that the error decreases very quickly as $\epsilon\propto \sqrt{2\pi p}e^{-p}$.

\par As an example, we use this MPO construction to simulate time evolution for the XXZ model in the thermodynamic limit,
\begin{equation*}
H_\mathrm{XXZ} =\sum_i S^x_iS^x_{i+1} + S^y_iS^y_{i+1} + \Delta S^z_iS^z_{i+1},
\end{equation*}
with $S^\alpha_i$ the spin-1/2 operators at site $i$ and we choose $\Delta=1/2$. This problem is closely related to the one considered in Ref.~\onlinecite{Trotzky2012} asserting the sumpremacy of quantum simulators. We use an MPO cluster expansion of the time-evolution operator with time step $dt$ up to clusters of size five, leading to a bond dimension of $1+4+16=21$ (see above), and we start our time evolution from the N\'{e}el state. We represent the time-evolved state as an MPS with bond dimension $\chi$. In each time step the bond dimension increases by applying the time-evolution MPO, and after a number of time steps we truncate the bond dimension down to $\chi$. We perform this truncation step by a variational optimization of the global overlap of the wavefunction, which can be done in a way very similar to other variational algorithms for uniform MPS \cite{Vanderstraeten2019, vomps}. 
\par
\begin{figure}
\subfigure{\includegraphics[width=\linewidth]{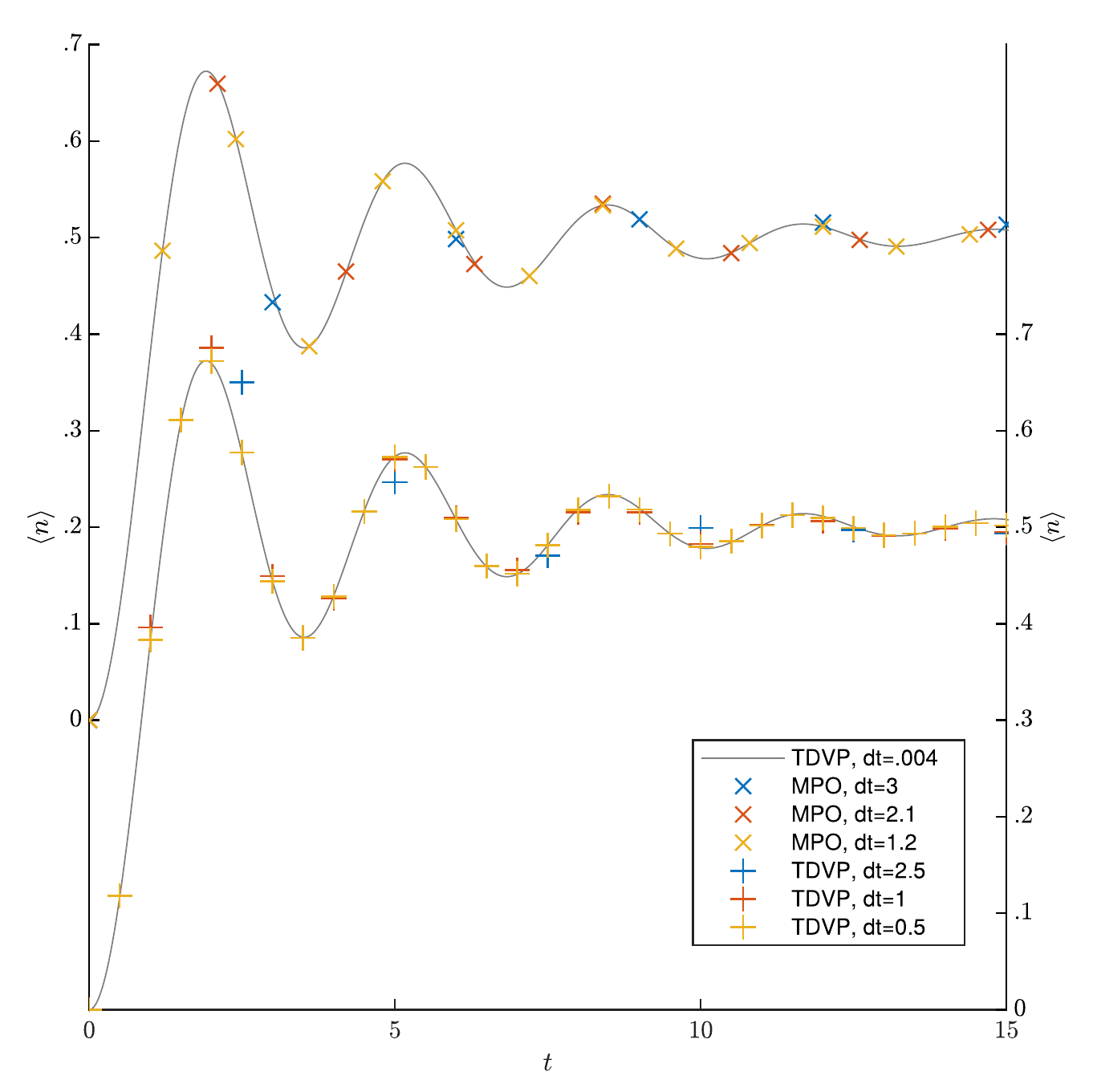}}
\subfigure{\includegraphics[width=\linewidth]{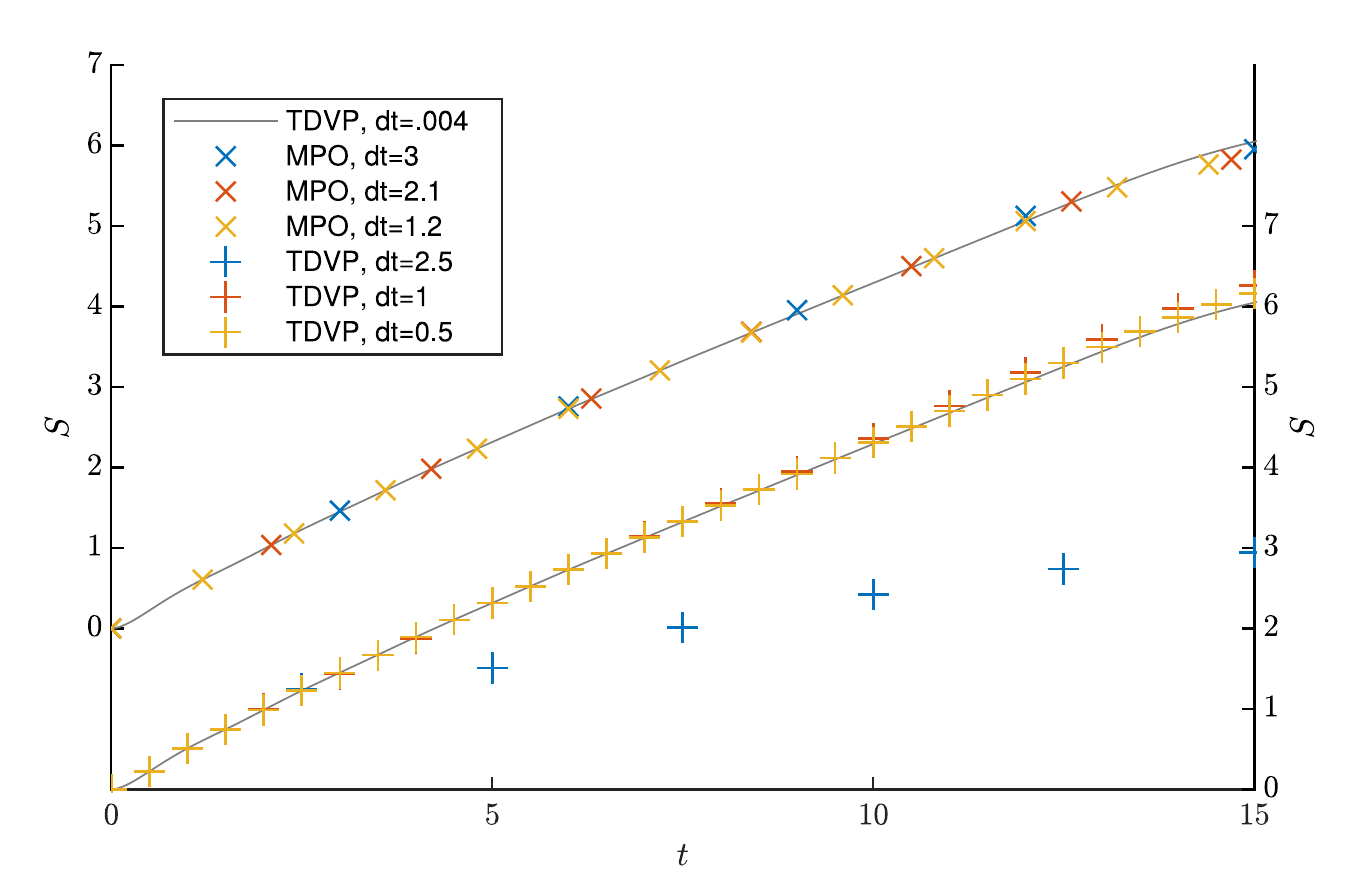}}
\caption{(Top) Time evolution of the occupation number for the N\'{e}el state evolved with the XXZ Hamiltonian with $\Delta=1/2$. We show results for different time steps for the MPO cluster expansion (crosses, left axis) and for TDVP evolution (pluses, right axis); both gray lines are the same reference result obtained with TDVP with very small time step. (Bottom) The same plot, now for the bipartite entanglement entropy. We have used $\mathrm{U}(1)$ symmetry in the MPO (where the largest subblock has dimension $D_\mathrm{max}=9$), and the largest block size of the time-evolved MPS is $\chi_\mathrm{max}=200$.}
\label{fig:evolution}
\end{figure}

 Our results are presented in Fig.~\ref{fig:evolution}. We have chosen to look at the occupation number per site as a function of time, $n=1/2+\braket{S^z}$, as well as the bipartite entanglement entropy $S$. We compare to simulations with the time-dependent variational principle (TDVP) for uniform MPS \cite{Haegeman2011}. We observe that the entropy grows linearly up to $t\approx13$, showing that up to this time the evolution is captured well by a finite-$\chi$ MPS. We see that the MPO cluster expansion produces almost exact results for time steps up to $dt=2.1$. Given that the error to the MPO goes as $dt^5$ this illustrates that the prefactor to this error is exceedingly small.

\renewcommand{\diagram}[1]{\,\vcenter{\hbox{\includegraphics[scale=0.3,page=#1]{./diagrams_pepo.pdf}}}\,}
\noindent\emph{Construction in two dimensions.}---%
The above construction for the MPO cluster expansion is generic, and can be applied to more general cases in higher dimensions. Here, we illustrate how the construction generalizes to the case of a \pepo representation for an exponentiated nearest-neighbour hamiltonian on a square lattice,
\begin{equation*}
\exp\left( t \sum_{\braket{ij}} h_{ij} \right) \approx \diagram{1}.
\end{equation*}
In a similar vein as for the MPO case, we incorporate the trivial part of the operator on the first virtual level and introduce a virtual level `$1$' for capturing the two-site clusters with the diagrams
\begin{equation*}
\diagram{2}, \quad  \diagram{3}, \quad \diagram{4}.
\end{equation*}
As in the 1D case, the three-site clusters, as well as some four- and five-site clusters, do not require a new virtual level, but can be encoded as
\begin{align*}
& \diagram{5},&  & \diagram{6}, \\ & \diagram{7}, & & \diagram{8}.
\end{align*}
\par In contrast to the one-dimensional case, loops will occur for large enough clusters, but this can be straightforwardly encorporated in the PEPO. The simplest loop involves the four-site cluster on a plaquette, which can be encoded by introducing a new virtual level `$2$',
\begin{equation*}
\diagram{9}.
\end{equation*}

\par Let us illustrate the power of this PEPO construction for PEPS ground-state optimization in the thermodynamic limit. The standard method for simulating imaginary-time evolution for PEPS is the full-update algorithm \cite{Jordan2008}, a method based on Trotter-Suzuki decompositions and local truncation steps. Here we approximate the imaginary-time evolution operator $\e^{-\tau H}$ with a PEPO cluster expansion and find its fixed point variationally using the algorithm in Ref. \onlinecite{Vanderstraeten2018}, in such as way that translational invariance and other global symmetries are conserved. In line with existing variational PEPS algorithms \cite{Corboz2016, Vanderstraeten2016}, the present PEPS algorithm performs a variational optimization of the imaginary-time evolution operator, and therefore the fixed point will converge as $\tau\to0$ to the optimal PEPS ground-state approximation for the Hamiltonian itself (in contrast to full-update optimization) \cite{Corboz2016, Vanderstraeten2016}. The computational cost and complexity of the algorithm is significantly smaller than the ones relying on variational optimization of the Hamiltonian itself \cite{Corboz2016, Vanderstraeten2016}.

\par We consider the Heisenberg Hamiltonian with sublattice rotation
\begin{equation*}
H = \sum_{\braket{ij}} -S^x_iS^x_j + S^y_iS^y_j -S^z_iS^z_j.
\end{equation*}
This particular sublattice rotation ensures that the PEPO is symmetric under reflection and complex conjugation, which simplifies the numerical optimization significantly \cite{vomps}. We have optimized PEPS with bond dimensions $D=3,4$ and different $\tau$'s, and listed the results for the energy in Table \ref{table}. We observe that we can reach the variational optimum as $\tau\to0$, but already obtain very precise values for the energy for a surprisingly large time step $\tau=0.1$.

\begin{table}[t]
\centering
\begin{tabular}{ | p{3cm} | p{2.2cm} | p{2cm} |}
\hline
& energy, $D=3$ & energy, $D=4$ \\ \hline
PEPO, $\tau=0.1$ & -0.6680688 & -0.6689614 \\ \hline
PEPO, $\tau=0.0562$ & -0.6680766 & -0.6689632 \\ \hline
PEPO, $\tau=0.0316$ & -0.6680792 &  -0.6689637 \\ \hline
PEPO, $\tau=0.0178$ & -0.6680806 &  -0.6689638 \\ \hline
PEPO, $\tau=0.01$ & -0.6680806  &   -0.6689638\\ \hline
variational PEPS\cite{Vanderstraeten2016} & -0.6680791 & -0.6689642 \\
\hline
\end{tabular}
\caption{Numerical results for the energy obtained with our variational PEPS algorithm based on the PEPO cluster expansion with imaginary-time step $\tau$; the PEPO has bond dimension five.}
\label{table}
\end{table}

\noindent\emph{Outlook.}---%
In this paper, we have explained how to construct cluster expansions for exponentiated local operators in quantum lattice systems with tensor networks. We have shown that these cluster expansions are of great practical use. On the one hand, they allow us to simulate time evolution of quantum spin systems with a much larger time step than in traditional approaches \cite{Paeckel2019}, and in such a way that symmetries are not broken. This is especially relevant for simulations starting from states with very few entanglement. The large time steps create entanglement very quickly, which has to be contrasted to schemes relying on the time-dependent variational principle \cite{Haegeman2011} which show problems in building up the entanglement. On the other hand, our cluster expansion provides a systematic scheme for converting the problem of finding ground states of Hamiltonians into one for finding the fixed point of a transfer matrix, which is a much better conditioned problem for two-dimensional problems. In addition, the PEPO inherits all spatial symmetries of the Hamiltonian and we do not introduce an artificial translational symmetry breaking. 
\par In this paper we have restricted to nearest-neighbour interactions, but our ideas can be applied to generic systems. In particular, it can be applied for simulating time evolution with MPS on a cylinder \cite{Zaletel2015}.
\par We expect that our PEPO construction will prove crucial for simulating dynamics with PEPS. On the one hand, one can deduce a PEPS real-time evolution scheme that does not rely on Trotter-Suzuki decompositions \cite{Czarnik2019, Hubig2019} for global quantum quenches without breaking any symmetries of the system, and, on the other, the computational cost of evaluating excitation energies with the quasiparticle ansatz \cite{Vanderstraeten2019b} can be drastically reduced. Finally, it is clear that the large step sizes enabled by the method are useful for constructing thermal states.

\noindent\emph{Acknowledgements.}---%
We thank Maarten Van Damme for running the TDVP with which we compared. We also thank Norbert Schuch for useful comments, and acknowledge support by the Research Foundation Flanders (LV) and the ERC grant QUTE (647905) (BV, FV).


%

\end{document}